%% file: main.tex
\documentclass{edm_article}

\usepackage{epstopdf}
\usepackage{caption}
\usepackage[labelformat=parens,labelfont={}]{subcaption}
\usepackage{comment}
\usepackage{algpseudocode}
\usepackage{multirow}
\usepackage{xcolor}
\usepackage{url}            

\newcommand{\norm}[1]{\left\lVert#1\right\rVert}

\DeclareMathOperator*{\argmin}{\arg\min}

\begin{document}

\title{DYNAMIC KNOWLEDGE EMBEDDING  AND TRACING}
%
\numberofauthors{2}
\author{
\alignauthor
Liangbei Xu\\
       \affaddr{Georgia Institute of Technology}\\
       \email{lxu66@gatech.edu}
\alignauthor
Mark A. Davenport\\
       \affaddr{Georgia Institute of Technology}\\
       \email{mdav@gatech.edu}
}

\maketitle


\begin{abstract}
  \input{Abstract}
\end{abstract}

%


\input{Introduction}

\input{Problem}

\input{MainResults}
\input{Conclusion}

\bibliographystyle{abbrv}
\bibliography{refs}

\end{document}

%% file: Abstract.tex
The goal of knowledge tracing is to track the state of a student's knowledge as it evolves over time. This plays a fundamental role in understanding the learning process and is a key task in the development of an intelligent tutoring system. In this paper we propose a novel approach to knowledge tracing that combines techniques from matrix factorization with recent progress in recurrent neural networks (RNNs) to effectively track the state of a student's knowledge. The proposed \emph{DynEmb} framework enables the tracking of student knowledge even without the concept/skill tag information that other knowledge tracing models require while simultaneously achieving superior performance.  We provide experimental evaluations demonstrating that DynEmb achieves improved performance compared to baselines and illustrating the robustness and effectiveness of the proposed framework. We also evaluate our approach using several real-world datasets showing that the proposed model outperforms the previous state-of-the-art. These results suggest that combining embedding models with sequential models such as RNNs is a promising new direction for knowledge tracing. 

%% file: Introduction.tex
\section{Introduction}
A central component in many computer-based learning systems, and in any kind of \emph{intelligent tutoring system} (ITS), is a method for estimating and tracking a student's knowledge or proficiency based on the student's previous interactions with the system. For example, a student may interact with many different course materials (homework exercises, quiz/exam questions, textbooks and other course materials, etc.) over a potentially long period of time. As a result of these interactions (as well as other external factors) the student's knowledge and proficiency will dynamically evolve over time~\cite{corbett1994knowledge,piech2015deep,cen2006learning,pavlik2009performance}. Tracking the state of a student's knowledge as it evolves can provide deeper understanding how the student is learning and which interactions (questions, textbooks, etc.) are most helpful, ultimately enabling the creation of a personalized learning environment tailored to provide an improved learning experience for the student. 

Estimating student knowledge or proficiency from a sequence of student interactions poses two fundamental challenges. First, student proficiency evolves over time as the student interacts with the system. For example, the student might turn to textbooks in response to getting a particular question wrong, and then may be able to answer a similar question correctly afterwards. Alternatively, the student may gradually lose proficiency in some areas if long periods of time pass without using this knowledge (e.g., over long vacations). Thus, we cannot treat this as a static problem of estimating a student's knowledge, but must think of this as a dynamic tracking problem.  A second and more subtle challenge is posed by the fact that the manner in which student proficiency evolves may be strongly influenced by the nature of the interactions.  For example, when a student is posed a question that requires knowledge of a particular concept, we not only learn something regarding the student's proficiency, but the student may also also learn something from the question.  In this way, the interactions both provide information to help us track the student's knowledge while simultaneously inducing changes in the state that we wish to track.

In this paper we propose a framework for tracing student knowledge using only a sequence of student responses to questions (for an ensemble of many students). The framework consists of two core components: a (static) embedding network that learns fixed latent representations of questions from student-question interactions and a recurrent neural network (RNN) that dynamically tracks the hidden state corresponding to each student's knowledge over time from the student's sequence of interactions. Our main contributions are:
\begin{itemize}
    \item A new knowledge tracing framework which exploits both the advantages of latent question embedding from response data and an RNN to track student knowledge;
    \item A framework that can track student knowledge without using the question-level concept/skill tags that other knowledge tracing models (e.g., DKT~\cite{piech2015deep} and its variants) require, avoiding labor-intensive manual tagging;
    \item A flexible framework that can also accommodate a variety of sequential modeling techniques (e.g., memory networks~\cite{zhang2017dynamic}) and can incorporate tag information and other features when available.
\end{itemize}

\section{Related work}
\subsection{Educational data mining}
Extracting useful information from the kind of educational data we consider was first studied within the \emph{intelligent tutoring} community. Since the seminal work of~\cite{corbett1994knowledge}, there has been a variety of efforts aimed towards understanding the cognitive processes that are most relevant in the context of an ITS, most of which aim to estimate students' proficiency based on their past interactions with the system with the aim of predict their performance on the new exercises/tests or customizing their learning materials.

\paragraph{Static models}
Item Response Theory (IRT) is a standard framework for modeling student responses to questions dating back to the 1950s~\cite{Linde_Handbook}. Perhaps the most common IRT model is the Rasch model~\cite{Rasch_Probabilisitc}. This is a simple two-parameter model in which each student is modelled as having a particular skill level and each question has a particular difficulty, which is then paired with a logistic link function to provide predictions of the probability a student will answer a question correctly.  There are natural mutlidimensional extensions of this and similar IRT models, which can be viewed as special cases of standard matrix factorization models (\cite{thai2010recommender}) or more general factorization machine models~\cite{rendle2010factorization}).

\paragraph{Sequential models}
Most of the models described above involve estimating a fixed student-question embedding which is then used to predict future responses. However, we fully expect the state of a student's knowledge to change over time.  To capture such dynamics, a natural approach is to incorporate dynamics in the model. One of the most popular models is Bayesian Knowledge Tracing (BKT), which employs a hidden Markov model (\cite{corbett1994knowledge}) to model the process of mastering a particular skill.
However, the BKT approach has some significant drawbacks. Most significantly, it models only a single skill or concept at a time. In practice, any particular question may be associated with a complex combination of different skills. To overcome this shortcoming, several alternative approaches have recently been proposed.

The most relevant attempt in this direction is the Deep Knowledge Tracing (DKT) framework~\cite{piech2015deep}. The DKT approach was inspired by recent progress in RNNs and deep RNN architectures. RNNs are a family of neural networks tailored for sequential prediction problems~\cite{williams1989learning}. In recent years deep RNN architectures have been shown to outperform many classical models in many application areas, including natural language processing and session-based recommendation system.  DKT is the first model to use RNNs to track student knowledge. DKT uses a one-hot encoding of skill/concept tags and associated responses as input and trains the RNN to predict the future student response. An extension of DKT is the Deep Hierarchical Knowledge Tracing (DHKT)~\cite{wang2019deep}, which extended DKT to incorporate problem IDs in addition to concept tags.

However empirical experiments in~\cite{xiong2016going,wilson2016back,wilson2016estimating} show that DKT does not appear to result in substantial improvement over many simpler models from classical IRT whose parameters and inferred states are psychologically meaningful.  It is worth noting that the IRT variants considered in~\cite{xiong2016going,wilson2016back,wilson2016estimating} use problem IDs as identifiers instead of skill IDs for DKT. Since multiple problem IDs can be tagged with the same skill IDs, we generally find that skill IDs repeat much more frequently than problem IDs.  Thus, a comparison using skill IDs would likely be more favorable to  a recurrent/sequential model like DKT. Of course, in considering only skill IDs we lose the ability to learn/exploit question-level information such as question difficulty.  Moreover, producing skill IDs for each question requires substantial human effort and is often not feasible in practice.  Furthermore all the experiments in~\cite{xiong2016going,wilson2016back,wilson2016estimating} consider the `New Student' evaluation protocol, which keeps a portion of the students as training sets and test on new students. Such an evaluation scenario may not be particularly meaningful in a real-world ITS and does not favor penalization models such as IRT, though online evaluation in~\cite{wilson2016back,wilson2016estimating} mitigates such bias. Thus, the comparison study in~\cite{xiong2016going,wilson2016back,wilson2016estimating} is not entirely satisfying and leaves open many questions regarding the potential benefits (or lack thereof) of deep RNNs for knowledge tracing.

\paragraph{Hybrid models}
There are also several attempts to combine static models and sequential models to exploit advantages from both approaches, such as the FAST model in~\cite{gonzalez2014general} and the LFKT model in~\cite{khajah2014integrating1}. In~\cite{khajah2014integrating2}, these two approaches are compared and the experimental results show that these two hybrid models do not outperform a simple IRT model. The authors conjecture that the lack of improvement is due to a confounding between item identity and the question position in a (nearly deterministic) sequence of questions. In contrast to these more pessimistic results, in this paper we propose a hybrid model and show that it can  harness the advantages from both static and sequential models in a way that outperforms both.

\subsection{Session-based recommendation systems}
A closely related application to knowledge tracing is that of predicting a user's preference for various items in a recommendation system. Among various recommendation systems, session based recommendation is the most closely related to knowledge tracing. For example, a session-based recommendation model, GRU4Rec, is proposed in~\cite{hidasi2015session} that has a similar architecture as DKT. However, GRU4Rec does not consider user identifications as inputs.  An alternative approach -- the Recurrent Recommender network (RRN)~\cite{wu2017recurrent} -- is capable of both modelling the seasonal evolution of items and tracking the user preferences over time. RRNs use a matrix factorization to model the stationary component of the user and item embeddings, and then two Long Short-Term Networks (LSTMs) to track the dynamic component of these embeddings. 

Though similar, there are some notable differences between product recommendation and knowledge tracing. First, user preferences tend to change much more slowly compared to student knowledge. Second, student interactions with questions have a significant impact on student knowledge, while in contrast interactions with an item (watching a movie, buying a product, etc.) typically have a mild impact at most on user preferences. Third, in a recommendation context, user responses may contain important implicit feedback~\cite{hu2008collaborative}. For example, we can conclude that a user will watch a movie or buy a product because he/she likes it, even if the user does not give explicit feedback. However, students typically have limited freedom to choose which questions to answer. These differences have important algorithmic implications.

%% file: Problem.tex
\section{The {DynEmb} framework}
\subsection{System architecture}
In this section we describe a novel framework for tracking student knowledge, dubbed \emph{DynEmb},  that learns a \emph{static} question embedding but exploits \emph{sequential} models of the temporal dynamics of student-question interactions to track the knowledge states of the students. We will represent our training data as a sequence of interactions of the form $\mathcal R_t = (s_t, q_t, r_t, o_t)$. Each interaction $\mathcal R_t$ involves a student $s_t$ and a question $q_t$. We assume there are $M$ questions and $N$ students. The response to the question is denoted $r_t$, which is most commonly a correct/incorrect binary outcome or occasionally a numerical score.  In this paper we focus mainly on the binary case, but the underlying framework can easily extend to the more general setting.  Finally, we let $o_t$ denote other information about the interaction that may be relevant, including -- but not limited to -- time stamps, questions tags, platform (e.g., paper, computer, mobile, etc.), and question text descriptions.

\begin{figure*}[tbp]
	\centering
	\includegraphics[width=0.80\linewidth]{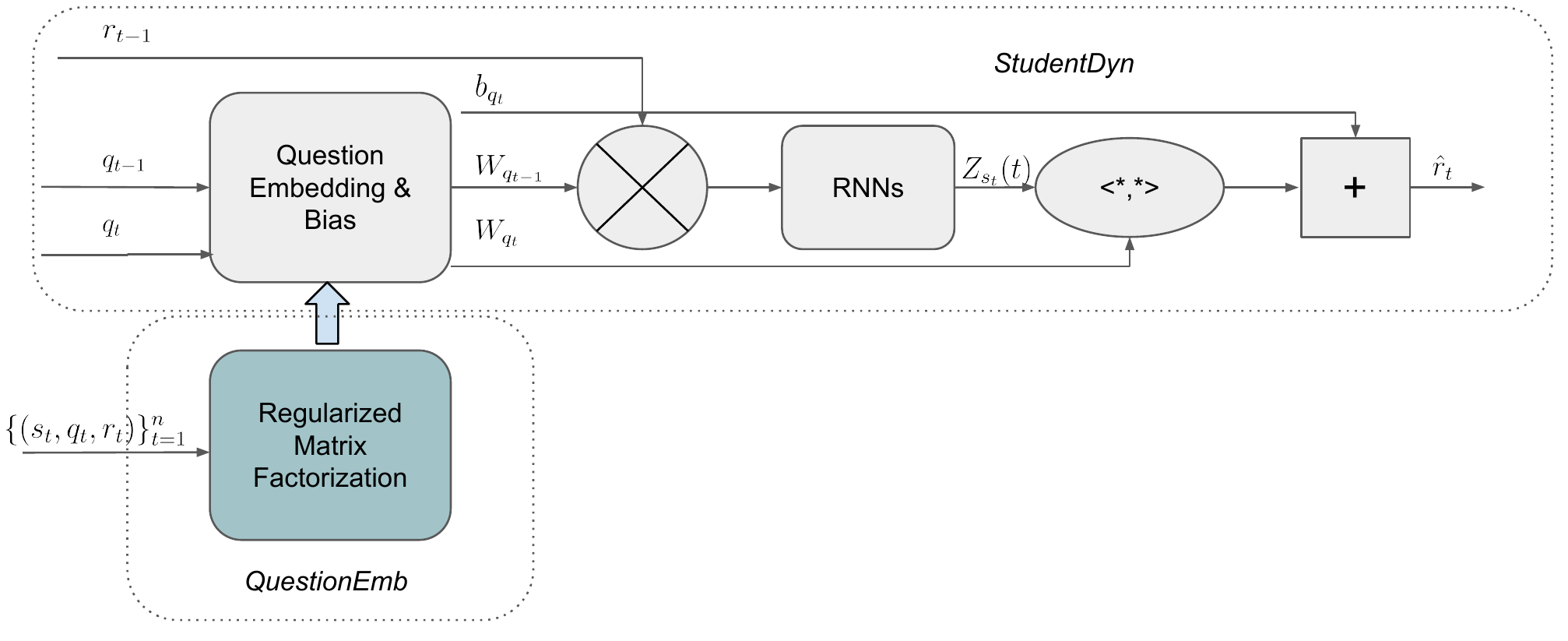} 
	\caption{Architecture for \emph{DynEmb}. First we train \emph{QuestionEmb} to obtain question embedding $W$ and bias $b$. Then we train the RNNs using past item embedding $W_{q_{t-1}}$ and response $r_{t-1}$ as inputs to track student knowledge.}
	\label{fig:architeture}
\end{figure*}

The goal of \emph{DynEmb} is to predict student responses to future questions given a historical sequence of interactions $\{\mathcal R_i\}_{i=1}^{n}$.  Specifically, given a new student-question pair $(s_t,q_t)$ and any additional information $o_t$ if available, our goal is to predict $r_t$. \emph{DynEmb} has two main components, each of which are trained independently (see~Figure \ref{fig:architeture}). The first component \emph{QuestionEmb} generates a $d-$dimensional \emph{question embedding} $W_{q_t}\in \mathbb R^d$ from $\{\mathcal R_i\}_{i=1}^{n}$ using standard matrix factorization techniques described in more detail below. The second component \emph{StudentDyn} learns to track each student's knowledge state using a sequential model that takes the student's past sequence of question embeddings $\{W_{q_i}\}_{i=1}^{t-1}$ and responses $\{r_i\}_{i=1}^{t-1}$ as inputs and produces a dynamic student embedding $Z_{s_t}(t) \in \mathbb R^{d}$.  The sequential model could be a ``vanilla'' RNN, a long short-term memory (LSTM) network, a gated recurrent unit (GRU), a memory network with attention, or others. In this work we use an LSTM in the \emph{StudentDyn} component by default. After obtaining the (static) question embedding $W_{q_t}$ and the (dynamic) student embedding $Z_{s_t}$, the predicted probability of a correct response is computed via
\begin{equation}
\label{eq:estimate-response}
    \hat r_t = \phi\left(\left\langle W_{q_t}, Z_{s_t}(t) \right\rangle  + b_{q_t}\right),
\end{equation}
where $b_{q_t}$ is a scalar that represents a bias learned for each question and $\phi$ is a sigmoid activation function. We describe these components in further detail below.

\paragraph{QuestionEmb} The \emph{QuestionEmb} component uses an $\ell_2$-regularized biased matrix factorization model to learn a static latent embedding for the  questions. More specifically, in this component we learn both a question embedding $W$ and a student embedding $Z$, where $W\in \mathbb R^{N\times d}$ is a matrix whose columns correspond to the question embedding vectors (the $W_{q}$'s) and $Z\in \mathbb R^{M\times d}$ is a matrix whose columns correspond to the student embedding vectors (the $Z_s$'s). These are learned via the following optimization problem: 
\begin{equation}
    \label{eq:matrix-factorization}
      \begin{split}
          \argmin_{W, Z, b, c}~&\sum_{t=1}^n\mathcal L \left( r_t, \phi\left(\left\langle W_{q_t}, Z_{s_t} \right\rangle  + b_{q_t} +  c_{s_t}\right) \right) \\
          & \quad + \lambda \left( \norm{W}_F^2 + \norm{Z}_F^2 \right),
      \end{split}
\end{equation}
where $b$ and $c$ are vectors of question and student ``biases'' respectively, $\lambda$ is the regularization parameter, and $\mathcal L(y, x) = -\left(y\log(x) + (1-y)\log(1-x) \right) $ is the log loss function. 
This is inspired by the observations in~\cite{xu2017simultaneous} that if the question embedding $W$ is static, then one can still use conventional matrix factorization to recover $W$, even though the other factors $Z$ may actually be changing over time. Finally, we note that while~\eqref{eq:matrix-factorization} is a non-convex optimization problem, simple optimization algorithms exist that provably converge to a global minimum~\cite{jain2013low,ge2016matrix}.

\paragraph{StudentDyn} The \emph{StudentDyn} component  uses an RNN to sequentially generate a student embedding after each interaction. For the case of a binary response, $r_{t-1}$, the input to the recurrent neural network is the Kronecker product of the question embedding learned by the \emph{QuestionEmb} component ($W_{q_{t-1}}$) and the vector $[r_{t-1}, 1-r_{t-1}]^T$. At time step $t$, an interaction between student $s_t$ and question $q_t$ is predicted via the model in~\eqref{eq:estimate-response}, and the RNN is trained to predict $r_t$.  The dynamic student embedding $Z_{s_t}(t)$ is the internal hidden state of the RNN, which is then combined with $W_{q_t}$ via~\eqref{eq:estimate-response} to obtain our final prediction.

\subsection{Model training} To train \emph{DynEmb}, we adopt a two-phase pretraining strategy. We first train the question embedding in the \emph{QuestionEmb} component. We then feed the learned question embedding to the \emph{StudentDyn} component to train the sequential model. Note that we keep the question embedding $W$ and the biases $b$ fixed when training the \emph{StudentDyn} component. This embedding pretraining strategy not only speeds up the training process, but also produces better prediction performance compared to end-to-end training (see Section~\ref{sec:pretrain} for an experimental justification). Similar pretraining strategies are widely used in learning complex models (e.g., for machine translation~\cite{qi2018and} and sentiment analysis~\cite{rezaeinia2017improving}).

Compared to DKT~\cite{piech2015deep}, DKVMN~\cite{zhang2017dynamic}, and other sequential knowledge tracing models, the explicit question embedding learned directly from interactions based on matrix factorization seems to be more robust. In fact, in our experiments we have observed that if we replace the (frequently repeating) concept/skill tags in DKT and DKVMN with the (much less frequently repeating) question identifiers, then both DKT and DKVMN will have significant performance degradation and require intensive computational resources to train. However, our model can track student knowledge using the pretrained question embedding instead of concept/skill tags. This allows our approach to exploit question difficulty information and scales well, especially when concept/skill tags are not available.

\subsection{Integrating skill tag information}
If manually-labeled skill tag information is available for each question, then it is convenient and beneficial to incorporate this information into the \emph{DynEmb} framework. However the question latent  space learned via the matrix factorization might be different from the latent space constructed by manual labeling. One simple method to exploit both approaches consists of concatenating the two latent question embeddings to form a new latent question embedding. The skill tags can be one-hot encoded. To further exploit the hierarchical relationship between questions and skill tags,  we initialize a question's embedding by the one-hot encoding of its corresponding skill tag, and put an additional $\ell_1$ regularization on the objective in~\eqref{eq:matrix-factorization} to promote sparsity.

To control the dimensionality of the latent space, the concatenated embedding is followed by a fully connected (FC) layer with ReLU activation.
This kind of integration scheme can be found in~\cite{cheng2016wide} and also enables easy incorporation of additional embeddings/fields, e.g., semantic embedding from question text. 

\begin{figure}[t]
	\centering
	\includegraphics[width=0.80\linewidth]{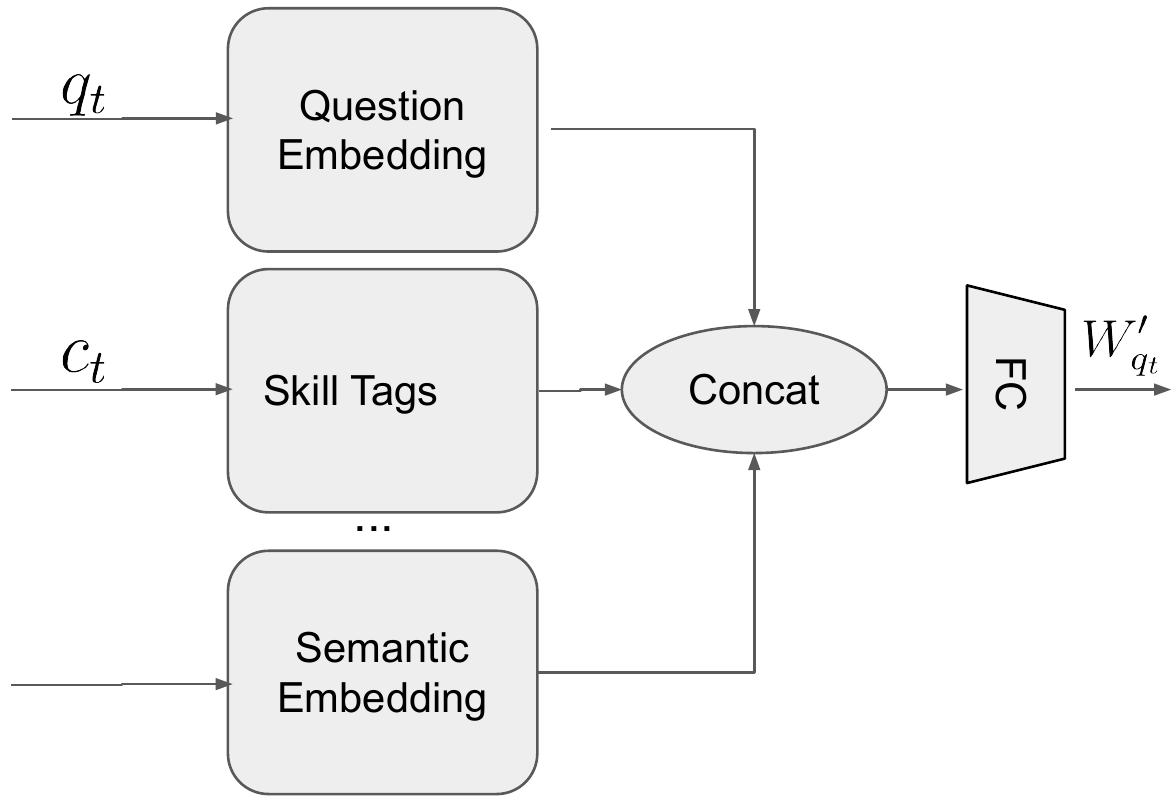} %
	\caption{Multiple input fields.}%
	\label{fig:multi-fields}%
\end{figure}

Finally, the \emph{StudentDyn} component uses an RNN to sequentially generate a student embedding after each interaction using this modified question embedding just as before. See~Figure~\ref{fig:multi-fields} for additional details.

%% file: MainResults.tex
\section{Experiments} 
\label{sec:kt-exp}
In this section, we experimentally validate the effectiveness of the proposed \emph{DynEmb} model on two tasks: prediction of response correctness for existing students and prediction of response correctness for new students. By conducting experiments on several data sets each and comparing with the relevant baselines, we show that:
\begin{enumerate}
    \item \emph{DynEmb} outperforms DKT by up to  5.43\% and 3.74\% in predicting the next response in the `New User'  and `Most Recent' evaluation settings respectively (see definition in Section~\ref{sec:kt-exp-setting});
    \item The performance of \emph{DynEmb} is stable with respect to the dimensionality of the item embedding;
    \item The proposed embedding pretraining strategy is a key component of the success of the \emph{DynEmb} approach.  
\end{enumerate}

\subsection{Experimental setting}
\label{sec:kt-exp-setting}
We consider the following baselines:
\begin{itemize}
    \item Algorithms that compute a static embedding: in this category, we compared with BMF~\cite{thai2010recommender}. We compare to both offline and online BMF. 
    \item Knowledge tracing based on RNNs: we compare with the state-of-the-art DKT algorithm~\cite{piech2015deep}.
\end{itemize}

We report the Area Under the ROC Curve (AUC) for comparing the predicted probabilities of correctness for each response. AUC is threshold agnostic, and is widely used in the knowledge tracing literature. 

We use two evaluation methods. The first is online response prediction for new users~\cite{piech2015deep, wilson2016back}. In this setting, students are first split into training and testing populations. Each model is first trained on the training population. Then for each time $t>1$ in each testing student's history, we train the student-level parameters in the model on a new student, including both the training population and the first $t-1$ interactions of the student history, computing the probability that the $t^{\text{th}}$ response is correct. In practice, we find that re-training and testing after each response is not computationally feasible for large datasets, in which case we perform online response prediction in batches. We denote this evaluation method the `New User' setting.  Our second method is to consider online response prediction for the the most recent interactions as in~\cite{wilson2016back}. The procedure here, denoted the `Most Recent' setting, is the same as in the `New User' setting except that we consider only the most recent interactions for our testing population as the testing data set.

\subsection{Experiment 1: Future response prediction}
In this experiment, the task is to predict students' response. The prediction task is: given all interactions up to time $t$, given the student $s$ and question $q$ involved in the interaction at time $t$, what is student $s$'s response (correct/incorrect) to question $q$?

We use the following data sets to evaluate performance on this task.

 \textbf{ASSISTments.} This data set was gathered from ASSISTments's skill builder problem sets, where students learn by working on similar questions until they can  respond correctly $n$ (usually 3) times in a row~\cite{assistmentsdata}. We use two one the provided data sets, ``ASSISTment09'' and ``ASSISTment12.'' Note that the authors updated ``ASSISTment09'' in 2017 (first found in~\cite{xiong2016going}).

 \textbf{Cognitive Tutor.} 
In the 2010 KDD Cup Challenge, the PSLC DataShop released several data sets from Carnegie Learning's Cognitive Tutor in (Pre-)Algebra from the years 2005-2009~\cite{kddcup}. We use three of the ``Development'' data sets, ``Algebra I 2005-2006,'' ``Algebra I 2006-2007,'' and ``Bridge to Algebra I 2006-2007.''

\begin{table*}[t]
\centering
\caption{Overview of data sets.}
\label{tab:my-table}
\setlength\tabcolsep{2pt}
\small
\begin{tabular}{|c|c|c|c|c|c|c|}
\hline
\multirow{2}{*}{Data set}        & \multicolumn{4}{c|}{Number of}           & \multirow{2}{*}{Ratio of correctness} & \multirow{2}{*}{Description} \\
                                 & Skills & Problems & Students & Responses &                                       &                              \\ \hline
\multirow{2}{*}{ASSISTments}     & 101    & 13111    & 4003     & 214424    & 0.658                                 & 2009                         \\ \cline{2-7} 
                                 & 265    & 47124    & 28998    & 2623624   & 0.699                                 & 2012                         \\ \hline
\multirow{3}{*}{Cognitive Tutor} & 90     & 210710   & 574      & 809693    & 0.767                                 & Algebra I 2005               \\ \cline{2-7} 
                                 & 488    & 580531   & 1338     & 2270384   & 0.772                                 & Algebra I 2006               \\ \cline{2-7} 
                                 & 494    & 207856   & 1146     & 3679188   & 0.888                                 & Bridge to Algebra 2006       \\ \hline
\end{tabular}
\end{table*}

\begin{table*}[t]
\centering
\setlength\tabcolsep{2pt}
\small
\caption{Future response prediction experiment: Table comparing the performance of \emph{DynEmb} (concatenating question and skill embedding) with baselines, in terms of AUC. \emph{DynEmb} outperforms the best baseline by up to 5.43\%. We also list the performance of \emph{DynEmb} with only question embeddings.}
\label{tab:results}
\begin{tabular}{|c|c|c|c|c|c|c|c|}
\hline
\multirow{2}{*}{Evaluation method} & \multirow{2}{*}{Model} & \multicolumn{2}{c|}{BMF} & \multirow{2}{*}{DKT} & \multicolumn{2}{c|}{DynEmb} & \multirow{2}{*}{Improvement} \\
                                   &                        & offline     & online     &                      & Question       & Concat      &                              \\ \hline
\multirow{5}{*}{New User}          & ASSISTment09           & 0.67        & 0.686      & 0.727                & 0.725         & 0.739       & 1.65\%                       \\ \cline{2-8} 
                                   & ASSISTment12           & 0.694       & 0.717      & 0.709                & 0.722         & 0.736       & 2.65\%                       \\ \cline{2-8} 
                                   & Algebra I 2005         & 0.761       & 0.763      & 0.773                & 0.803         & 0.815       & 5.43\%                       \\ \cline{2-8} 
                                   & Algebra I 2006         & 0.761       & 0.786      & 0.808                & 0.805         & 0.821       & 1.61\%                       \\ \cline{2-8} 
                                   & Bridge to Algebra 2006 & 0.838       & 0.844      & 0.856                & 0.868         & 0.873       & 1.99\%                       \\ \hline
\multirow{5}{*}{Most Recent}       & ASSISTment09           & 0.706       & 0.727      & 0.661                & 0.738         & 0.727       & 0.00\%                       \\ \cline{2-8} 
                                   & ASSISTment12           & 0.67        & 0.696      & 0.71                 & 0.692         & 0.714       & 0.56\%                       \\ \cline{2-8} 
                                   & Algebra I 2005         & 0.744       & 0.763      & 0.779                & 0.791         & 0.808       & 3.72\%                       \\ \cline{2-8} 
                                   & Algebra I 2006         & 0.761       & 0.782      & 0.801                & 0.813         & 0.822       & 2.62\%                       \\ \cline{2-8} 
                                   & Bridge to Algebra 2006 & 0.831       & 0.839      & 0.847                & 0.859         & 0.865       & 2.13\%                       \\ \hline
\end{tabular}
\end{table*}

\paragraph{Preprocessing of data sets} As noted in~\cite{wilson2016back}, there are multiple records duplicating a single interaction (represented by a unique \emph{order\_{id}} value) in ``ASSISTment09.'' These duplicate rows arise when a single interaction is aligned with multiple skills. This provides DKT models access to the ground truth when making their predictions, which can artificially boost prediction results by a significant amount. We adopt two strategies to clean the data. The first is to discard rows duplicating a single interaction (as in~\cite{wilson2016back}); the second is to combine these duplicating rows into a single row with a new skill tag as suggested by~\cite{xiong2016going}. In this paper we removed duplicate and multiple-skill repeated records in all data sets to ensure fairness for the purpose of comparison. We also removed ``not original'' records as suggested by~\cite{xiong2016going}. 
We do similar cleaning operation on the other data set ``ASSISTment12''. 
For the Cognitive Tutor data sets, we form problem identifiers from the concatenation of the ``Problem Name'' and ``Step Name'' fields. 

\paragraph{Implementation details} The dimensionality of the input to the RNNs in \emph{DynEmb} is fixed at $100$. The $\ell_2$ regularization parameter in the \emph{QuestionEmb} component is chosen using cross-validation  based on standard BMF. The hyper-parameters in the \emph{StudentDyn} component are the same as DKT and chosen by cross-validation. 

\paragraph{Results}  Table~\ref{tab:results} compares the results of \emph{DynEmb} with the baseline. We observe that \emph{DynEmb} significantly outperforms the best baseline in all datasets in terms of AUC on the three datasets up to 5.43\%.

\subsection{Experiment 2: Robustness to embedding dimensionality}
\begin{figure}[htp]
	\centering
	\includegraphics[width=0.90\linewidth]{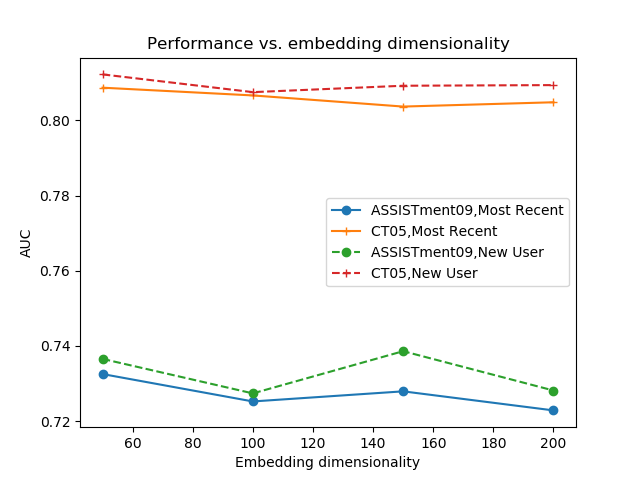} %
	\caption{Performance versus embedding dimensionality. }%
	\label{fig:embedding_size}
\end{figure}

In this section, we study the effect of the dynamic embedding dimensionality on the tracking performance.
In this study we use the ``ASSISTment09'' and Cognitive Tutor ``Algebra I 2005'' (``CT05'' for short) datasets, which have the smallest number of interactions from the two tutoring systems respectively. The effect on other datasets
is similar and omitted for the sake of brevity. We will test on the response prediction task. As we can see from Figure~\ref{fig:embedding_size}, the performance by AUC of \emph{DynEmb} is quite stable over a wide range of embedding dimensionalities. This robustness is an additional attractive feature of our approach.  

\subsection{Experiment 3: Embedding pretraining vs. end-to-end training}
\label{sec:pretrain}
\begin{figure}[htp]
	\centering
	\includegraphics[width=0.90\linewidth]{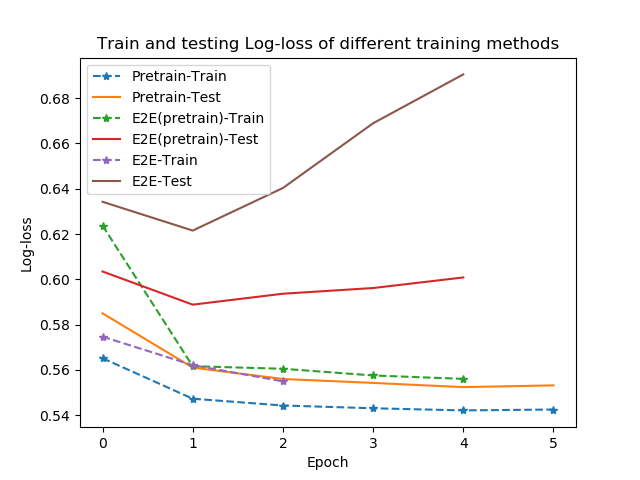} %
	\caption{Training and testing log-loss of different training methods. }%
	\label{fig:why_pretrain}
\end{figure}

In this section we demonstrate why \emph{DynEmb} uses pretraining for the question embedding. The dataset used in this section is ``ASSISTment09.'' We use the ``Most Recent'' evaluation method. In Figure~\ref{fig:why_pretrain}, we can see that end-to-end (E2E for short) training (with/without pretraining the question embedding) will cause over-fitting, while the learning curve of proposed pretraining strategy does not suffer from over-fitting or under-fitting. Of course, another advantage of pretraining is its improved computational efficiency. The combination of these two factors provides powerful evidence for choosing pretraining over an end-to-end training strategy in this framework.

\subsection{Experiment 4: Visualizing question embedding}
\begin{figure}[htp]
	\centering
	\includegraphics[width=0.90\linewidth]{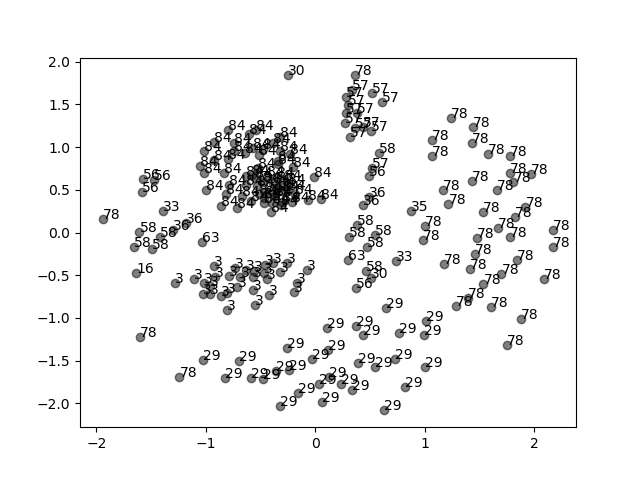} %
	\caption{Visualization of the embedding of random selection of 200 questions by multidimensional scaling. }%
	\label{fig:emb_viz}
\end{figure}

Though the latent space of the question embedding learned via matrix factorization is not explicitly aligned  with the latent space formed by the manually-labeled skill tags that were provided, the proposed question embedding initialization and sparsity promotion is remarkably effective at aligning the question embedding space with the manually constructed skill embedding space. This provides additional semantic meaning for the learned question embedding, which improves model interpretability.   Figure~\ref{fig:emb_viz} shows clear clustering of question embedding with respect to the associated skills (indicated by skill identifiers). 

%% file: Conclusion.tex
\section{Conclusion and discussion}
In this paper we presented a framework to track student knowledge in an ITS by utilizing techniques from matrix factorization/embedding and RNNs. Our framework can track student knowledge without the concept/skill tag information required by other knowledge tracing models, e.g., DKT~\cite{piech2015deep} and its variants. This avoids labor-intensive manual tagging. Taking advantage of additional latent question embeddings, our framework outperforms recent state of the art knowledge tracing models using RNNs. By constructing an embedding of the questions via matrix factorization in addition to skill tags, our framework can fuse question-level and skill-level information. The \emph{DynEmb} framework is also flexible in that it can accommodate various matrix factorization techniques and dynamical models, which makes it a promising avenue for future research and development of algorithms for knowledge tracing. 

However, in the context of a  real-world implementation, several challenges remain regarding how to design a practical \emph{DynEmb} based system for knowledge tracing. For example, developing a method amenable to deployment in an online setting will require additional algorithmic improvement. Another challenge concerns how to incorporate additional sources of auxiliary information not considered here, such as question text or details about additional student interactions with an ITS (browsing history, textbook interactions, etc.)  to best exploit all of the information that might be available. We believe that the DynEmb framework provides a natural platform to address such challenges.

%% file: main.bbl
\begin{thebibliography}{10}

\bibitem{cen2006learning}
H.~Cen, K.~Koedinger, and B.~Junker.
\newblock Learning factors analysis--a general method for cognitive model
  evaluation and improvement.
\newblock In {\em International Conference on Intelligent Tutoring Systems},
  pages 164--175. Springer, 2006.

\bibitem{cheng2016wide}
H.-T. Cheng, L.~Koc, J.~Harmsen, T.~Shaked, T.~Chandra, H.~Aradhye,
  G.~Anderson, G.~Corrado, W.~Chai, M.~Ispir, et~al.
\newblock Wide \& deep learning for recommender systems.
\newblock In {\em Proceedings of the 1st workshop on deep learning for
  recommender systems}, pages 7--10. ACM, 2016.

\bibitem{corbett1994knowledge}
A.~T. Corbett and J.~R. Anderson.
\newblock Knowledge tracing: Modeling the acquisition of procedural knowledge.
\newblock {\em User modeling and user-adapted interaction}, 4(4):253--278,
  1994.

\bibitem{ge2016matrix}
R.~Ge, J.~D. Lee, and T.~Ma.
\newblock Matrix completion has no spurious local minimum.
\newblock In {\em Advances in Neural Information Processing Systems}, pages
  2973--2981, 2016.

\bibitem{gonzalez2014general}
J.~Gonz{\'a}lez-Brenes, Y.~Huang, and P.~Brusilovsky.
\newblock General features in knowledge tracing to model multiple subskills,
  temporal item response theory, and expert knowledge.
\newblock In {\em The 7th International Conference on Educational Data Mining},
  pages 84--91. University of Pittsburgh, 2014.

\bibitem{hidasi2015session}
B.~Hidasi, A.~Karatzoglou, L.~Baltrunas, and D.~Tikk.
\newblock Session-based recommendations with recurrent neural networks.
\newblock {\em arXiv preprint arXiv:1511.06939}, 2015.

\bibitem{hu2008collaborative}
Y.~Hu, Y.~Koren, and C.~Volinsky.
\newblock Collaborative filtering for implicit feedback datasets.
\newblock In {\em 2008 Eighth IEEE International Conference on Data Mining},
  pages 263--272. Ieee, 2008.

\bibitem{jain2013low}
P.~Jain, P.~Netrapalli, and S.~Sanghavi.
\newblock Low-rank matrix completion using alternating minimization.
\newblock In {\em Proceedings of the forty-fifth annual ACM symposium on Theory
  of computing}, pages 665--674. ACM, 2013.

\bibitem{khajah2014integrating1}
M.~Khajah, R.~Wing, R.~Lindsey, and M.~Mozer.
\newblock Integrating latent-factor and knowledge-tracing models to predict
  individual differences in learning.
\newblock In {\em Educational Data Mining 2014}. Citeseer, 2014.

\bibitem{khajah2014integrating2}
M.~M. Khajah, Y.~Huang, J.~P. Gonz{\'a}lez-Brenes, M.~C. Mozer, and
  P.~Brusilovsky.
\newblock Integrating knowledge tracing and item response theory: A tale of two
  frameworks.
\newblock In {\em CEUR Workshop Proceedings}, volume 1181, pages 7--15.
  University of Pittsburgh, 2014.

\bibitem{assistmentsdata}
Z.~Pardos.
\newblock Assistments dataset homepage.
\newblock \url{https://sites.google.com/site/assistmentsdata/home/}.

\bibitem{pavlik2009performance}
P.~I. Pavlik~Jr, H.~Cen, and K.~R. Koedinger.
\newblock Performance factors analysis--a new alternative to knowledge tracing.
\newblock {\em Online Submission}, 2009.

\bibitem{piech2015deep}
C.~Piech, J.~Bassen, J.~Huang, S.~Ganguli, M.~Sahami, L.~J. Guibas, and
  J.~Sohl-Dickstein.
\newblock Deep knowledge tracing.
\newblock In {\em Advances in neural information processing systems}, pages
  505--513, 2015.

\bibitem{qi2018and}
Y.~Qi, D.~S. Sachan, M.~Felix, S.~J. Padmanabhan, and G.~Neubig.
\newblock When and why are pre-trained word embeddings useful for neural
  machine translation?
\newblock {\em arXiv preprint arXiv:1804.06323}, 2018.

\bibitem{Rasch_Probabilisitc}
G.~Rasch.
\newblock {\em Probabilistic Models for Some Intelligence and Attainment
  Tests}.
\newblock Danish Institute for Educational Research, Copenhagen, Denmark, 1960.

\bibitem{rendle2010factorization}
S.~Rendle.
\newblock Factorization machines.
\newblock In {\em 2010 IEEE International Conference on Data Mining}, pages
  995--1000. IEEE, 2010.

\bibitem{rezaeinia2017improving}
S.~M. Rezaeinia, A.~Ghodsi, and R.~Rahmani.
\newblock Improving the accuracy of pre-trained word embeddings for sentiment
  analysis.
\newblock {\em arXiv preprint arXiv:1711.08609}, 2017.

\bibitem{kddcup}
J.~Stamper, A.~Niculescu-mizil, S.~Ritter, G.~G.J~Gordon, and K.~Koedinger.
\newblock Challedge data sets from kdd cup 2010.
\newblock \url{https://pslcdatashop.web.cmu.edu/KDDCup/downloads.jsp}.

\bibitem{thai2010recommender}
N.~Thai-Nghe, L.~Drumond, A.~Krohn-Grimberghe, and L.~Schmidt-Thieme.
\newblock Recommender system for predicting student performance.
\newblock {\em Procedia Computer Science}, 1(2):2811--2819, 2010.

\bibitem{Linde_Handbook}
W.~van~der Linden and R.~Hambleton, editors.
\newblock {\em Handbook of Modern Item Reponse Theory}.
\newblock Springer-Verlag, New York, NY, 2010.

\bibitem{wang2019deep}
T.~Wang, F.~Ma, and J.~Gao.
\newblock Deep hierarchical knowledge tracing.
\newblock In {\em The 12th International Conference on Educational Data
  Mining}, pages 671--674. University of Buffalo, 2019.

\bibitem{williams1989learning}
R.~J. Williams and D.~Zipser.
\newblock A learning algorithm for continually running fully recurrent neural
  networks.
\newblock {\em Neural computation}, 1(2):270--280, 1989.

\bibitem{wilson2016back}
K.~H. Wilson, Y.~Karklin, B.~Han, and C.~Ekanadham.
\newblock Back to the basics: Bayesian extensions of irt outperform neural
  networks for proficiency estimation.
\newblock {\em arXiv preprint arXiv:1604.02336}, 2016.

\bibitem{wilson2016estimating}
K.~H. Wilson, X.~Xiong, M.~Khajah, R.~V. Lindsey, S.~Zhao, Y.~Karklin, E.~G.
  Van~Inwegen, B.~Han, C.~Ekanadham, J.~E. Beck, et~al.
\newblock Estimating student proficiency: Deep learning is not the panacea.
\newblock In {\em In Neural Information Processing Systems, Workshop on Machine
  Learning for Education}, page~3, 2016.

\bibitem{wu2017recurrent}
C.-Y. Wu, A.~Ahmed, A.~Beutel, A.~J. Smola, and H.~Jing.
\newblock Recurrent recommender networks.
\newblock In {\em Proceedings of the tenth ACM international conference on web
  search and data mining}, pages 495--503. ACM, 2017.

\bibitem{xiong2016going}
X.~Xiong, S.~Zhao, E.~G. Van~Inwegen, and J.~E. Beck.
\newblock Going deeper with deep knowledge tracing.
\newblock {\em International Educational Data Mining Society}, 2016.

\bibitem{xu2017simultaneous}
L.~Xu and M.~A. Davenport.
\newblock Simultaneous recovery of a series of low-rank matrices by locally
  weighted matrix smoothing.
\newblock In {\em 2017 IEEE 7th International Workshop on Computational
  Advances in Multi-Sensor Adaptive Processing (CAMSAP)}, pages 1--5. IEEE,
  2017.

\bibitem{zhang2017dynamic}
J.~Zhang, X.~Shi, I.~King, and D.-Y. Yeung.
\newblock Dynamic key-value memory networks for knowledge tracing.
\newblock In {\em Proceedings of the 26th international conference on World
  Wide Web}, pages 765--774. International World Wide Web Conferences Steering
  Committee, 2017.

\end{thebibliography}
